\documentclass[preprint,showpacs,preprintnumbers,amsmath,amssymb,superscriptaddress]{revtex4}
\usepackage{graphicx} % Include figure files
\usepackage{dcolumn}  % Align table columns on decimal point
\usepackage{bm}       % bold math
\begin{document}
\preprint{}
\title{Origin and residence time of water in the Lima Aquifer}
\author{M. Montoya$^{1,2}$ and E. Mamani$^2$\\
$^1$~Instituto Peruano de Energ\'ia Nuclear, Canad\'a 1470, SB, Lima, Per\'u\\
$^2$Universidad Nacional de Ingenier\'ia, T\'upac Amaru 210, R\'imac, Lima, Per\'u.\\}

\begin{abstract}
 The 8 million inhabitants of the coast Lima City are supplied with water from Rimac and Chillons rivers and water wells in the Lima aquifer. Historics of Rimac River flow and static level of water level in wells are correlated in order to calculate residence time of water since the aquifer is recharged by Rimac River until water reaches a well located 12 km farther, in Miraflores district near sea. Relative abundances of ${}^{2}$H and ${}^{18}$O are used to identify origins of waters from those wells. ${}^{3}$H and ${}^{14}$C contents, respectively, are used to estimate ages of waters.
\end{abstract}

\maketitle
\section{Introduction}

\noindent The purpose of this paper is to study the evolution of the Lima aquifer dynamics. This aquifer is exploited by the Lima Waterworks Company (SEDAPAL) for its water to supply the growing Lima population, which in 1997 reached up to 8 million inhabitants. 

\noindent In 1992, a 24.8 m${}^{3}$/s demand for water in Lima was confronted. This demand was satisfied with surface water of the Rimac river (9-13m${}^{3}$/s), the aquifer (up to 9.5 m${}^{3}$/s) replenished by the Rimac and Chillon rivers. A 2.2 m${}^{3}$/s overexploitation with increasing trend was observed.

\noindent In 1997, the exploited amount of groundwater reached 12.38 m$^3$/s. That year projects for a more rational use of groundwater, artificial Rimac. aquifer recharge, and switching groundwater to surface water began \cite{quintana2002}. 

\noindent In order to evaluate the effects of those projects, SEDAPAL measured the evolution of the static level of its water wells. To study the origin of the waters in the aquifer, the relative abundance of ${}^{2}$H and ${}^{16}$O were measured. The residence time in the aquifer of waters is estimated from contents of  ${}^{3}$H.  

\section{The Rimac Basin}

\noindent Located along the west coast of South America, from 3 degrees to 18 degrees south latitude, Peru has three distinguishable regions called Costa, Sierra and Selva, respectively. The Costa is the coast limited by the Pacific Ocean; the Sierra, is a part of the Andean mountains, some of them glaciers above 6000m; and the Selva is the west part of the tropical Amazonas basin. The rainy season on the Sierra, from December to Mars, results in significant fresh water flow in the rivers, most of them tributary to the Amazon River and the rest of them flowing down to the Pacific Ocean on a few drainage basins. Lima City is on the central part of the coast. It is irrigated mainly by the Rimac River, whose slope is more than 3\%. The Rimac River flows from the wetlands, small lakes and glacial meltwaters of the Cordillera Central through steep narrow valleys onto a clastic wedge of coarse alluvial sediments between the mountains and the coast ~\cite{leavell2008}.

\noindent The Rimac basin has a length of 204 km, an average wide of 16 km and a surface of 3398 km${}^{2}$. Its borders are the Mantaro Basin to the North-east, Lurin Basin to the South-east, Chillon Basin to the North-west and the Pacific Ocean to the South-west ~\cite{ingemmet1988}.
The Rimac River drops from 5000 m above the sea level flowing down 145 km before to discharge into the sea. Its main tributary is the Santa Eulalia River. 

\subsection{ Climates, Rainfall and Temperature}

\noindent 

\noindent Due to the cold Humboldt Current, Lima's climate is cooler than cities of the same latitude and altitude in other parts of the world. Its precipitation is 21 mm annually. At the altitudes from 1000 m to 2000 m above sea level, since December ti April, there is a low intensity rainy season.  At the altitudes from 2000 to 3500 m, the precipitations are 250 mm annually and occur from December to March. From 3500 m to 4000 m, it rains from December to March, and the annual rainfall is over 450 mm.  Over 4000 m, the rainfall can reach 750 mm annually, and the night temperatures is average are 4C average ~\cite{ingemmet1988}.

\noindent 

\subsection{Geological characteristics}

\noindent 

\noindent The Rimac basin geological structures are composed by sedimentary, metamorphic, volcanic and intrusive rocks of Jurassic and the Quaternary ages. One can distinguish folds, faults and plutonic and hypabyssal masses. Two geological zones are identified: the Occidental Zone, formed by with bodies of igneous, sedimentary and metamorphic rocks; and the Oriental Zone formed by rocks of Jurassic, cretaceous, tertiary and quaternary ages ~\cite{ingemmet1988}.

\noindent  

\subsection{Hydrological description} 

\noindent 

\noindent The Rimac Hydrographic Basin covers an area of 3398 km${}^{2}$, having an average gradient of 3.23 ~\cite{ingemmet1988}, with 2237.2 km${}^{2}$ with permanent precipitations and 895.2 km${}^{2}$ with intermittent precipitations ~\cite{cesel1999}. The Rimac River has 23 tributaries, the main of which is the Santa Eulalia basin with an area of 1097.7 km${}^{2}$ ~\cite{cesel1999}.  The top lines of the Rimac´s tributaries are between 4400m to 5200m ~\cite{rojas1994}.

\noindent 

\noindent The dry lower zone of the Rimac Basin is formed by the Lima's entrance to its mouth in the Pacific. This zone is 17.5 km long with a gradient of 1.1 \% and altitudes from 195 m to the sea level ~\cite{rojas1994}. 

\noindent 

\section{Hydrogeological description of Lima aquifer}

\noindent 

\noindent The Lima aquifer is formed by unconsolidated alluvial, interspersed layers of gravels, sands, silt, and mudstones, deposited over a low permeability material, bounded by volcanic- sedimentary rocks and granites in the substrate. The area of the Lima aquifer is 260 km${}^{2}$, with a thickness estimated to be between 400 m to 500m ~\cite{ingemmet1988}. 

\noindent The Rimac and Chillon rivers recharge the Lima aquifer through filtrations, garden watering and irrigation canals. The sub-surface flows that begin in the upper parts of the Rimac and Chillon basins also contribute to the water table of Lima. These flows reach the lower levels of the water table on the ends of the bay.  In the sea, in front of Callao, there is a well that is being pumped to supply water to ships. In Chorrillos there is another water well whose floor is below sea level. 

\noindent The permeability in the valley is 1 x 10${}^{-3}$ m/s and changes to 10${}^{-4}$ m/s in the alluvial cone. The storage coefficient is 5 \% in the valley and 0.2 \% in the coastal area ~\cite{ingemmet1988}. 

\noindent 

\noindent The upper part of Lima aquifer is mainly composed by almost 100 m of gravels and other coarse-grained sediments in sand and clay matrix interspersed with fine-grained layers. The lower part is formed by much finer unconsolidated sediments composed of sands, silts and mud. The greater part of the aquifer is mostly unconsolidated alluvial deposits. Between the Rimac and Chillón rivers, the upper part is formed by fine-grained deposits ~\cite{rojas1994, mendez2005}. 

\noindent 

\section{Lima Delta} 

\noindent 

\noindent On the surface of the Lima aquifer, a delta with the shape of an equilateral triangle is identified. The triangle is formed by a) 20 km of the Rimac River, which contributes to the aquifer replenishing, and flows from East to West at latitude -12${}^\circ$; b)  the Surco River, which begins from the latitude -76.90 ${}^\circ$, and flows in north-east to south-west direction; and c) the beach (Pacific Ocean), where the  aquifer discharges. See Fig. 1. 

\noindent The north part of the Lima Delta also receives contribution from the Chillon River. The East end of the Rimac River in the delta has an altitude of 300 m above sea level. In the middle section of the northern side of the Lima delta, the Rimac alluvial deposits meet with the ones that correspond to the Chillon alluvial deposits. The ground level, at that meeting point, is at 130 m above sea level, and consequently the topographic level decreases in both sides of the Rimac River. This topographical surface suggests that the Rimac alluvial deposits descended from east to west and then dispersed to both, north and south sides. In north direction, 8km from Rimac River, the ground level decreases to 66 m above sea level. The ground level retrieves the 130m altitude at 25km north from the Rimac River. Thus, near Chillon river, the alluvial deposits from Rimac basin are at higher levels than those from Chillon basin.

\noindent We have taken water samples from 25 wells and 2 water springs owned by SEDAPL whose locations are numerated in Fig. 1. On this Fig. the geographical coordinates of these wells and water springs are indicated.  These coordinates are represented as (Longitude, Latitude). The beach forms a bay with a length of about 25km, between the water well called La Punta (1), with geographical coordinates (-77.16${}^\circ$, -12.06${}^\circ$), and the water well of Chorrillos (21), with geographical coordinates (-77.01${}^\circ$, -12.17${}^\circ$). In a beach close to Chorrillos, known as Costa Verde, a cliff with a height of 60m is formed. At the base of these cliffs, by the sea, the Estrella water spring (19), with coordinates (-77.02${}^\circ$, -12.14${}^\circ$), and the Barranquito water spring (20), with coordinates (-77.01${}^\circ$, -12.17), are found. 

\noindent

\subsection{Residence Time of Water in the Lima Aquifer}

\noindent

\noindent The Rimac River, which flows from east to west, replenishes the northern part of the water table over the Lima Delta. There is a well located 1 km northeast from the Estrella spring. This water well, numbered as 71 by SEDAPAL, had a static level of -33m in 1966 and - 58 m in 1990, level at which it stabilized, probably because the wells of the Lima aquifer did not reach to extract water at this level. 

\noindent Rimac river flow reached a maximun historic level in 1986, which was reflected in 1989 as a maximum in the historic of static level of well 71. Similarly, in 2001, the flow of the Rimac reached another maximum level, which resulted in a maximum in the static level of the well 71 in 2004. See Fig. 2. Annual variations of Rimac river flow in the period 2000-2006 are reflected in the annual variations of the static level of well 71 in the period 2003 - 2009. See Fig. ~\ref{fig:fig3}. This suggests that the water infiltrated from Rimac River in the water table under the Lima delta takes around 3 years to reach the Estrella water spring near sea.

\noindent

\subsection{Origin of waters}

\noindent

\noindent Origin of water is determined with the natural tracers ${}^{2}$H and ${}^{18}$O, which are stable isotopes. The composition of the waters of these isotopes are expressed in terms of ($\delta $ ${}^{18}$ O, $\delta $ ${}^{2}$H), which are expressed in units of .

\noindent The Lima aquifer receives contributions of Chillon and Rimac basins. SEDAPAL has built water wells in both basins, at higher altitudes than the Lima delta, before the waters are mixed. The geographical coordinates of SEDAPAL water wells are shown in Fig. 1. The corresponding isotopic composition of water from these wells is shown in Fig. 4. The isotopic composition of the Chillon and Rimac basins, are clearly distinguishable as different columns on the $\delta $-diagram. 

\noindent In contrast, in the aquifer under the Lima Delta, the water seems to come from both basins, except those on the opposite sides of the delta, which maintain their corresponding identities.

\noindent The Estrella (19) and the Barranquito (20) water springs have the isotopic compositions (-13.85, -102.58) and (-13.56, -102. 32), respectively. The positions of these values ??on the $\delta $-diagram (Fig. 4), suggest that the Estrella water spring is a mix of waters from Rimac and Chillon river basins, while the Barranquito water spring only contains water from Rimac basin.

\noindent The composition of water from Pueblo Libre water well (16) is (-14.02, -103.63), which resembles the parameters that characterize the waters from the Chillon basin. See Fig. 4. The water well of Callao Sea (2) has a composition of (-13.94, -102.53) which resembles the parameters of the water from the Rimac basin. On the other side, the composition of the water from Puente Piedra (6) is (-12.03, -90.76), which makes it the heaviest among the studied water samples.

\noindent Water weight affects the vaporization cycle of water: lighter water falls with rain at higher elevations than heavier ones. This means that among the samples taken, the waters of Pueblo Libre and Callao fell at the highest altitude and the water from Puente Piedra (6) well fell at a the lowest altitude. Notice that Puente Piedra (6) well is located on the right side of the Chillon River, separated from most of the samples that have been taken from the left side of the river.

\noindent

\subsection{Water age}

\noindent

\noindent Let's define as the well´s floor altitude, the altitude of the well minus the well´s depth. In Fig. 5, the tritium content depending on the altitude of the well´s floor is plotted. One can observe that, whenever the altitude is higher than sea level, the tritium content increases with the well´s floor altitude.

\noindent It is worth noting that the water of the well under the sea in front of Callao is 0.1 TU. If one assumes that in fresh water the tritium abundance is 2 TU, then the water from this well has been around 50 years underground. The waters of La Punta and Chorrillos wells have 0.4 and 0.5 TU, respectively, suggesting an age of around 25 years.

\noindent The low ${}^{14}$C content in the waters of the mentioned wells (see Fig. ~\ref{fig:fig6}) seems to confirm the hypothesis that these waters belong to the oldest water samples.

\noindent

\subsection{Oxigen Content}

\noindent

\noindent In Fig. ~\ref{fig:fig7}, the relative abundance of ${}^{18}$O in water as a function of the well´s floors altitude of water wells located in Lima is plotted. Most wells in the Lima delta have their floors below sea level. The abundances of ${}^{18}$O in water from the samples of the delta Lima are between -14.02 and -13.12 . The sample from Pueblo Libre well, whose altitude is -112.9m, has -14.02 of $\delta $${}^{ 18}$O, which is the lowest value obtained from all the samples. The water sample from the Callao well, whose floor altitude is -15m, contains -14.04 of $\delta $${}^{ 18}$O. This means that these wells have the lightest waters, suggesting that they belong to rains that fell in the highest parts of the Rimac and Chillon basins, respectively.

\noindent The waters of the two wells of Ate Vitarte (26,27), whose floors are above sea level, have similar low levels of ${}^{18}$O, which makes them exceptions among wells with high altitudes. One of these wells, whose floor is at 166.2 m above sea level, contains -13.92  of $\delta $ ${}^{18}$O. The other well, whose floor is at 281.1 m above sea level, contains -13.86  of $\delta {}^{18}$O. This suggests that these waters belong to the same aquifer layer whose waters reach Callao and Pueblo Libre wells, whose corresponding floors are below sea level.

\noindent The contents of ${}^{2}$H and ${}^{18}$O of La Molina well (25) are separate from the Lima Delta. Their values belong to ?? upper layers of the water table.

\noindent As it was shown in Fig. ~\ref{fig:fig7}, the water from well Puente Piedra (6), whose floor is at 131.55 m above sea level, contains -12.03  of $\delta $ ${}^{18}$O, which corresponds to the heaviest water among the collected samples.

\noindent

\section{Conclusions}

\noindent

\noindent The Lima aquifer is formed by several alluvial layers that fall in direction East to West in the Rimac basin, and in direction North to South in Chillon basin.

\noindent The relative abundance of ${}^{16}$O and ${}^{2}$H,  expressed as ($\delta $ ${}^{18}$ O, $\delta $ ${}^{2}$H) in units of , in water samples from wells in Lima aquifer are distinguishable as two columns on the $\delta $-diagram. Using those isotopes as tracers, the ($\delta $ ${}^{18}$ O, $\delta $ ${}^{2}$H) values in the water samples from Lima Delta aquifer (south of Rimac River) suggest a contribution from the Rimac and Chillon basins, with a preponderance of contribution of Chillon in the northern part, and a preponderance of contribution of Rimac basin in the southern part of the Lima aquifer.  

\noindent The contents of ${}^{18}$O, in function to the altitude of the well´s floors, allow distinguishing between the Rimac and the Chillon origin of waters. This differentiation is higher in $\delta {}^{18}$O measurements, depending on the altitude above sea level of the well´s floor.

\noindent The evolution of the static levels of Lima Delta (South of Lima) water wells and ${}^{3}$H contents suggest that the recharges of the aquifer by the Rimac River have a residence time of about 3 years until it reaches the sea.

\noindent The ${}^{3}$H content in waters from wells whose floors have lower altitudes than sea level, located in the opposite ends of the beach, had decades of residence time. The well located at the northwest of Lima Delta has a water of about 30 years old, and the water of the well located at the southwest of the delta has 20 years old.

\noindent

\acknowledgments{We thank to International Atomic Energy Agency (IAEA) for ${}^{3}$H and ${}^{14}$C contents measuring; to SEDAPAL for the wells static level data.  Stimulating discussions with Ruben Rojas (IPEN), Edgar Alva (SEDAPAL), Fluquer Peña (INGEMMET), David Wahl and Carolina Aguilar are gratefully acknowledged.}

\newpage

\begin{figure}%[!ht]
\centering
\includegraphics[width=8cm]{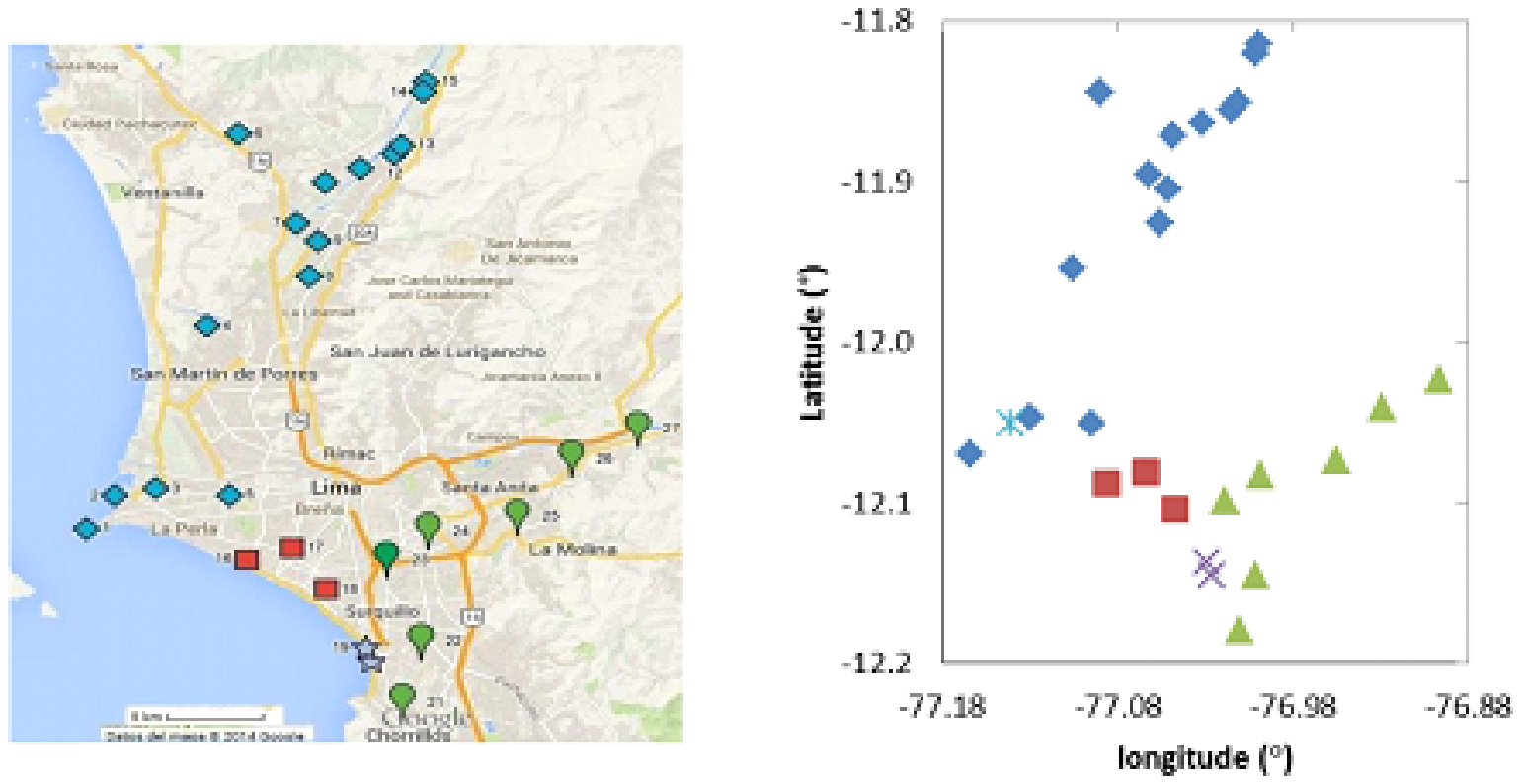}
\caption{Satelite image of Lima City. One can define the Lima Delta as an triangle with the northern side formed by the Rimac River flowing from East to West, crossing Ate, Lima and discharging on Sea in front of Callao. The Rimac River remplishes the aquifer of Lima Delta at the north part of Lima.  The alluvial deposits from Rimac basin fell from East to West, and are dispersed as well to the North as to the South of Rimac river; while the deposits from Chillon basin fell roughly from North to the South (left). Geographical coordinates of a group of wells used by the Lima Waterworks Company (SEDAPAL). The diamonds correspond to wells located mostly in the basin of Chillon. The triangles correspond to the Rimac basin. The squares correspond to waters supplied from both basins. The asterisk refers to a well in the seabed. The blades represent two water springs on the beach of Costa Verde. A well in the sea is identified by an asterisk (right)}
\label{fig:fig1}
\end{figure}

\noindent 

\begin{figure}%[!ht]
\centering
\includegraphics[width=8cm]{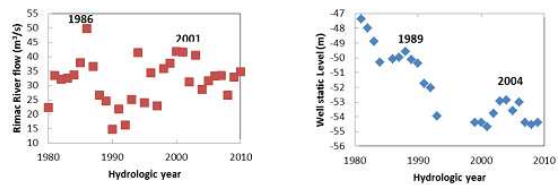}
\caption{Historic flow (m${}^{3}$/s) between 1980 and 2010 of the Rimac River in Chosica, 50 km east from the sea. It is observed a maximum in 1986 and another one in 2001 (left). Historic of the static level of well 71, 1 km from the Miraflores beach. The graph shows two maxima in 1989 and 2004. Both maxima occurred three years after the two corresponding maxima in the flow of the Rimac river (right)}
\label{fig:fig2}
\end{figure}

\noindent 

\begin{figure}%[!ht]
\centering
\includegraphics[width=8cm]{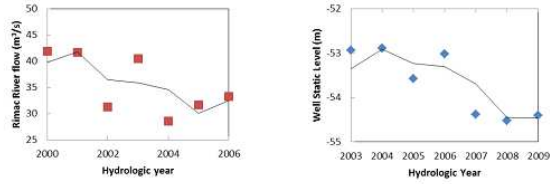}
\caption{(Left): the same as Fig. ~\ref{fig:fig2} (left) referred to period 2000 -- 2006. (right): the same as Fig. ~\ref{fig:fig2} (right), corresponding to period 2003 -- 2009. One can observe similar fluctuations of static level with a 3 years delay referred to flow fluctuations}
\label{fig:fig3}
\end{figure}

\noindent 

\begin{figure}%[!ht]
\centering
\includegraphics[width=8cm]{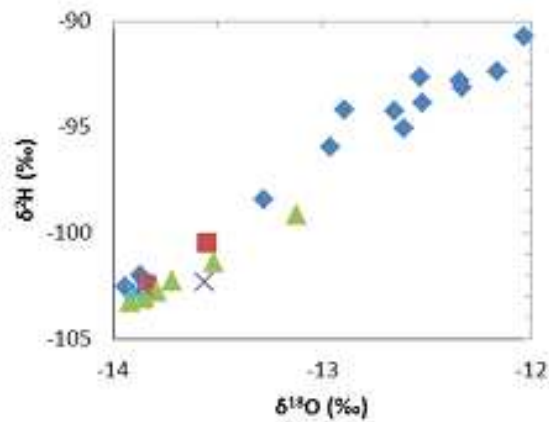}
\caption{ $\delta $-diagram. Isotopic relative abundance in $\delta {}^{18}$O and $\delta {}^{2}$H of water corresponding to wells presented in Fig. ~\ref{fig:fig2}. The highest content of those isotopes correspond to sample from the well of Puente Piedra 1, while the lowest content corresponds to one well of Callao, close to sea}
\label{fig:fig4}
\end{figure} 

\noindent 

\begin{figure}%[!ht]
\centering
\includegraphics[width=8cm]{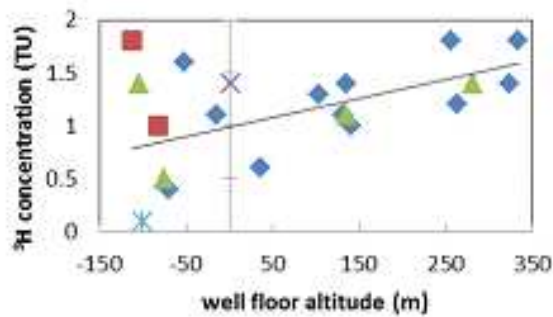}
\caption{ Tritium concentration in water sample as a function of the altitude of water well floor corresponding to wells represented in Fig.~\ref{fig:fig2}. For altitudes over sea level the tritium content increases lineally with water well floor altitude}
\label{fig:fig5}
\end{figure}

\noindent 

\begin{figure}%[!ht]
\centering
\includegraphics[width=8cm]{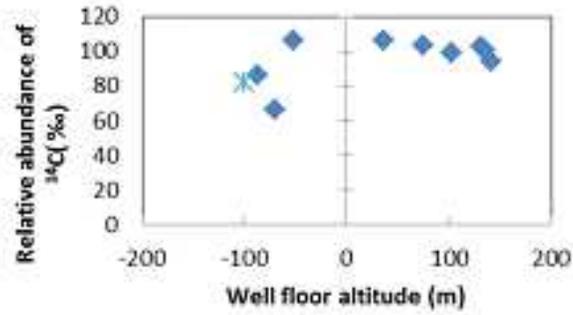}
\caption{Relative abundance of ${}^{14}$C as a function of water well floor altitude referred to some wells represented in Fig. ~\ref{fig:fig2}.}
\label{fig:fig6}
\end{figure} 

\noindent 

\begin{figure}%[!ht]
\centering
\includegraphics[width=8cm]{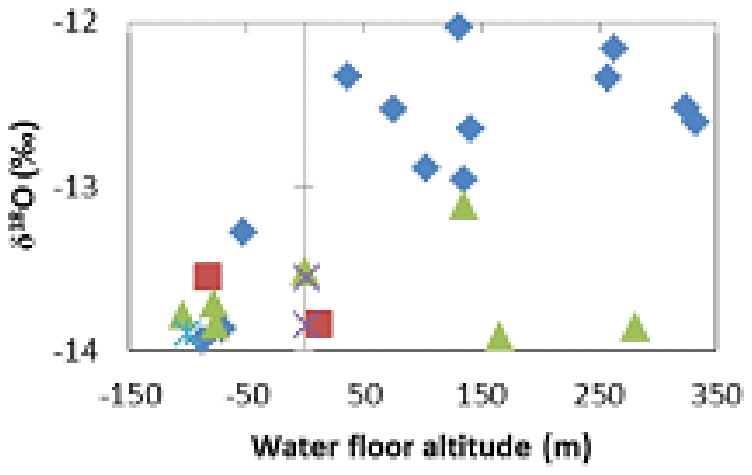}
\caption{ Relative abundance of ${}^{18}$O in water samples as a function of altitude of the water well floor, referred to water wells represented in Fig. ~\ref{fig:fig2}}
\label{fig:fig7}
\end{figure} 


\noindent

\begin{thebibliography}{}

\bibitem{quintana2002}  Quintana, J. and Tovar, J. (2002) Evaluaci\'on del acu\'ifero de Lima (Per\'u) y medidas correctoras para contrarrestar la sobreexplotaci\'on. Boletín Geol\'ogico y Minero, 113 (3): 303-312 

\bibitem{leavell2008} Daniel N. (2008) The Consequences of Climate Change for the Water Resources of Per\'u. Proceedings, XIII World Water Congress, International Water Resources Association, Montpelier, FR. (Proceedings paper on CD).

\bibitem{ingemmet1988} INGEMMET (1988) Estudio Geodin\'amico de la Cuenca del Río Rimac. (Bolet\'in Nº 8b Serie C). Instituto Geol\'ogico Minero y Metalúrgico. Lima. 263 p.

\bibitem{cesel1999} CESEL (1999) Monitoreo Complementario de la Cuenca del Río Rimac. Dirección General de Asuntos Ambientales, Ministerio de Energ\'ia y Minas, Lima. 39 s.

\bibitem{rojas1994} Rojas, R., Howard, G. and Bastram, J. (1994) Groundwater quality and water supply in Lima Peru. Groundwater Quality. Chapman and Hall, London UK.123 p.

\bibitem{mendez2005} M\'endez, W. (2005) Contamination of Rimac River Basin Peru, due to Mining Tailings. MSc Thesis. TRITA-LWR Master Thesis LRW-EX-05-23. Lima, Peru. 31 p Ann.

\end{thebibliography}
\end{document}